\documentclass[5p,twocolumn]{elsarticle}

\usepackage{amssymb}

\usepackage{lineno}

\journal{Nuclear Instruments and Methods A}

\begin{document}

\begin{frontmatter}



\title{X-ray radiographic technique for measuring density uniformity of silica aerogel}


\author[First,Second]{Makoto Tabata\corref{cor1}}
\ead{makoto@hepburn.s.chiba-u.ac.jp}
\cortext[cor1]{Corresponding author.} 
\author[Third]{Yoshikiyo Hatakeyama}
\author[Fourth]{Ichiro Adachi}
\author[Third]{Takeshi Morita}
\author[Third]{Keiko Nishikawa}

\address[First]{Institute of Space and Astronautical Science (ISAS), Japan Aerospace Exploration Agency (JAXA), Sagamihara, Japan}
\address[Second]{Department of Physics, Chiba University, Chiba, Japan}
\address[Third]{Graduate School of Advanced Integration Science, Chiba University, Chiba, Japan}
\address[Fourth]{Institute of Particle and Nuclear Studies (IPNS), High Energy Accelerator Research Organization (KEK), Tsukuba, Japan}

\begin{abstract}
This paper proposes a new X-ray radiographic technique for measuring density uniformity of silica aerogels used as radiator in proximity-focusing ring-imaging Cherenkov detectors. To obtain high performance in a large-area detector, a key characteristic of radiator is the density (i.e. refractive index) uniformity of an individual aerogel monolith. At a refractive index of $n$ = 1.05, our requirement for the refractive index uniformity in the transverse plane direction of an aerogel tile is $|\delta (n - 1)/(n - 1)| < 4\%$ in a focusing dual layer radiator (with different refractive indices) scheme. We applied the radiographic technique to evaluate the density uniformity of our original aerogels from a trial production and that of Panasonic products (SP-50) as a reference, and to confirm they have sufficient density uniformity within $\pm $1\% along the transverse plane direction. The measurement results show that the proposed technique can quantitatively estimate the density uniformity of aerogels.
\end{abstract}


\end{frontmatter}


\section{Introduction}
\label{}
In high-energy and nuclear experiments, silica aerogel, which is a colloidal form of quartz (SiO$_2$), is important as a radiator in Cherenkov counters. This is because this transparent solid has a tunable refractive index ($n$) and it is easier to handle than liquid or gas radiators. Hydrophobic aerogels with $n$ = 1.006--1.14 can be consistently manufactured by our conventional method \cite{cite1}. Furthermore, by using a new method (the pin-drying method), we can produce ultrahigh-refractive-index ($n$ = 1.10--1.26) aerogels with sufficient transparency \cite{cite2}. The recent progress toward developing highly transparent aerogels with $n$ = 1.04--1.26 \cite{cite3} will open a new window to particle identification (PID) in future experiments, for example, the super $B$-factory experiment at KEK (Belle II experiment \cite{cite4}) and nuclear experiments proposed at J-PARC.

Our aerogel development is mainly motivated by the Belle detector upgrade program, including the PID device upgrade. For the new PID device to separate kaons from pions in the momentum range around 1--4 GeV/$c$ in the forward end-cap of the Belle II detector, the A-RICH group is developing a proximity-focusing aerogel ring-imaging Cherenkov (A-RICH) detector \cite{cite5}. In a dual layer radiator scheme (focusing combination) \cite{cite6}, the use of approximately 250 large-area aerogel tiles with $n \sim $ 1.05 and dimensions 18 $\times $ 18 $\times $ 2 cm$^3$ is planned. Section 2 describes the maximum allowed refractive index variation over a large transversal area that meets our requirement to ensure a constant-mean single-photon Cherenkov angle distribution.

On the basis of scanning electron microscopy, the typical diameter scale of the silica primary particles and structural pores in aerogels is known to be of the order of 10 nm. In the micrometre range, aerogels are considered to be sufficiently uniform for optical use. However, it is not self-evident that our aerogels are uniform enough in a macroscopic range from millimeters to centimeters. Local non-uniformities might be introduced during all the phases of aerogel production (i.e. wet-gel synthesis, ageing, hydrophobic treatment and supercritical drying). Therefore, in this paper, we propose a method to evaluate the density uniformity of an aerogel monolith using X-rays, and we demonstrate the validity of the method for our standard aerogels as a scale model of the A-RICH radiator.

\section{Importance of refractive index uniformity}
\label{}

\subsection{Influence of refractive index uniformity on RICH detectors}
In a RICH detector, it is crucial to accurately determine the refractive index over the entire tile of an individual aerogel because the single-photon Cherenkov angle resolution, in addition to the mean of the single-photon Cherenkov angle distribution, is affected by the effective local refractive index of the radiator. The refractive index uniformity of an individual aerogel monolith is generally a key property of a good Cherenkov radiator, similar to aerogel transparency, because of the impact on detector performance. The refractive index uniformity can be studied along the thickness and transverse plane directions.

Uniformity along the thickness direction is more important because non-uniformities together with chromatic dispersion of the refractive index would directly worsen the Cherenkov angle resolution. In the dual layer radiator scheme, the aerogel of each layer is expected to be sufficiently uniform or to have a limited refractive index gradient \cite{cite7} along the thickness direction. In previous beam tests \cite{cite6}, \cite{cite8}, \cite{cite9}, the prototype proximity-focusing RICH counters were composed of two radiator tiles each 20 mm thick and a photon detector array parallel to the radiator face with an expansion distance of 20 cm. As a rule, charged particles impacted on a limited area at the center of the aerogel tiles. Therefore, uniformity in the thickness direction, rather than in the transverse plane direction, contributes most to the Cherenkov angle resolution. With this configuration, the Cherenkov angle resolution $\sigma _\theta$ $\sim $ 14 mrad was obtained. The result agrees with those of an analytical model \cite{cite10} assuming $\sigma _{rest}$ = 6 mrad, where $\sigma _{rest}$ is the contribution to the error in determination of the Cherenkov angle from sources other than the emission point uncertainty and due to the position resolution of the detector (see Ref. \cite{cite10} for more detail). This contribution from other sources may be partially explained by non-uniformity along the thickness direction. This suggests a fruitful direction for future investigation.

As a first step, we now focus on uniformity in the transverse direction, which is the main subject of this study. The measurement of the refractive index uniformity in the transverse plane direction is mandatory for the A-RICH detector because it has a large area of 3.5 m$^2$. To guarantee a constant-mean single-photon Cherenkov angle distribution over an entire aerogel tile, it is crucial to evaluate uniformity along the transverse plane direction. Our final goal is to measure the uniformity of large-area aerogel tiles with dimensions of 18 $\times $ 18 $\times $ 2 cm$^3$ produced by both conventional and pin-drying methods. Measurements of uniformity will provide a comprehensive evaluation that includes the contribution of uniformity in the thickness direction. The maximum tolerance of the refractive index non-uniformity is $|\delta (n - 1)/(n - 1)| < 4\%$, which corresponds to $n$ = 1.05 $\pm $ 0.002. This acceptable variation is based on the model calculation in Ref. \cite{cite10}, where the degradation of the Cherenkov angle resolution in the dual layer radiator scheme is very small. This was empirically verified in a dedicated beam test \cite{cite11}.

\subsection{Previous methods}
A direct method for measuring the refractive index uniformity of aerogels by a laser (the gradient method) was first introduced in Ref. \cite{cite12}, and was also applied in Refs. \cite{cite13}, \cite{cite14}, \cite{cite15}. This method is based on the deviation of the laser due to the transverse gradient of the refractive index along its optical path. If aerogels have a variation in thickness or a bend (i.e. surface tilt or meniscus effects), this method will not be suitable for measuring the refractive index uniformity because the tilted surface refracts the laser even if the aerogel is sufficiently uniform. In Ref. \cite{cite15}, the thickness variation within an aerogel was mapped using a mechanical comparator, and the result of the gradient method was corrected by the effects of thickness variation. The thickness variation will affect the optical characteristics of aerogels and the RICH detector performance. In Ref. \cite{cite16}, the thickness variation was evaluated using a laser-based sensor to investigate aerogels for use in a RICH radiator.

Another method for evaluating the refractive index uniformity of aerogels by X-rays was applied in Ref. \cite{cite17}. The first study in which X-rays were used to examine density variations in aerogels for use in a RICH radiator was presented in Ref. \cite{cite18} using a digital radiographic device \cite{cite19}. In Ref. \cite{cite18}, X-ray images of an aerogel block (top and side views) were used. A multiple-layer aerogel block was also measured by X-rays in a later study \cite{cite20}.

Cherenkov imaging using a RICH detector is the most thorough method for evaluating the local refractive index of aerogels in a beam test. In Ref. \cite{cite21}, the variation in the Cherenkov angle over an aerogel surface was measured by focusing the beam impact position on the aerogel; however, the measurement area was limited to 36 segments of 5 $\times $ 5 mm$^2$ each.

\begin{table*}[t]
\centering 
\caption{Aerogel specifications measured in this study.}
\label{table:table1}
	\begin{tabular}{lcccc}
		\hline
		ID (our reference) & $n$ * & $\Lambda _T$ ** [mm] & Dimensions [mm$^3$] & Density [g/cm$^3$] \\
		\hline
		Panasonic1 (XRA14) & 1.0514 & 16 & 115 $\times $ 115 $\times $ 11.1 & 0.195 $\pm $ 0.004 \\
		Panasonic2 (XRA15) & 1.0516 & 17 & 115 $\times $ 115 $\times $ 11.1 & 0.194 $\pm $ 0.004 \\
		Chiba (BTR3-3a) & 1.0501 & 42 & 93 $\times $ 93 $\times $ 20.0 & 0.175 $\pm $ 0.002 \\
		\hline
		\multicolumn{5}{l} {*Refractive index at 405 nm. **Transmission length at 400 nm.} \\
	\end{tabular}
\end{table*}

\subsection{Proposal for novel X-ray radiographic technique}
Here we present an X-ray radiographic technique as another method to investigate the density uniformity within an individual aerogel monolith. In this technique, X-ray absorption in the aerogel material is measured more quantitatively. By conducting an experimental procedure as described in the next section, we can determine the absolute density of local areas of aerogels. The refractive index of aerogels is proportional to their density $\rho $, $n-1 = k\rho $, where $k$ is a wavelength-dependent coefficient. This is an empirical (Lorentz--Lorenz based \cite{cite27}) relationship \cite{cite1}. Measuring the density uniformity is thus a means of evaluating the refractive index uniformity of an individual aerogel tile. A characteristic of our method is that it measures the aerogel thickness as accurately as possible. Thickness measurements have important implications for the quantitative evaluation of density of aerogels because the X-ray absorption is sensitive to changes in their thickness. Thus, we expose the cross-sectional surface by cutting aerogels in order to measure the thickness. The potential disadvantage is that our technique destroys the aerogels. However, by measuring the aerogel thickness precisely, our X-ray radiographic technique has an important advantage because the measurement of the density of aerogels does not depend on the particular condition of their surface. A preliminary experimental result for our aerogel using the X-ray technique was briefly reported in Ref. \cite{cite22}. In the following pages, a complete investigation of the density uniformity of basic (conventional) aerogels with small dimensions is reported. For novel aerogels produced by the pin-drying method, the results of characterization studies are presented in a separate paper \cite{cite23}.

\section{X-ray radiographic technique}
\label{}

\subsection{Measurement concept}
X-ray absorption by materials obeys the exponential attenuation law
\begin{equation}
\label{eq:eq1}
I/I_0=\exp(-\mu _mx),
\end{equation}
where $I_0$ is the incident X-ray intensity, $I$ is the transmitted X-ray intensity, $\mu _m$ is the X-ray mass absorption coefficient and $x$ is the mass thickness of the material. This equation is valid for a narrow monoenergetic X-ray beam. The mass thickness is defined as
\begin{equation}
\label{eq:eq2}
x=\rho t,
\end{equation}
where $t$ is the thickness of the material. The photon mass absorption coefficients for elements are available as a function of the photon energy from the database in Ref. \cite{cite24}. We found that the material density is reproduced well when we use the database. In the present study, the photon energy for X-ray absorption measurements was 8.04 keV. The mass absorption coefficients of compounds and mixtures can be obtained as follows:
\begin{equation}
\label{eq:eq3}
\mu _m=\sum_{i}w_i(\mu _m)_{i},
\end{equation}
where $w_i$ is the fraction by weight of the $i$th atomic constituent. The density of materials is described using Eqs. (\ref{eq:eq1}) and (\ref{eq:eq2}) as follows:
\begin{equation}
\label{eq:eq4}
\rho =\frac{1}{\mu _mt}\ln{\frac{I_0}{I}}.
\end{equation}
To calculate the density of aerogels using Eq. (\ref{eq:eq4}), we need to measure the mass absorption coefficient, the X-ray transmittance and the thickness of aerogels. Therefore, the evaluation of the density uniformity of aerogels consists of the following three steps:
\begin{itemize}
\item elemental ratio analysis by X-ray fluorescence (XRF);
\item X-ray absorption measurement;
\item thickness measurement.
\end{itemize}

\subsection{Aerogel samples}
In this study, three hydrophobic aerogels with a refractive index of 1.05 were analyzed. Two of them were provided by Panasonic (Matsushita) Electric Works, Ltd. (products: SP-50). The two aerogels, obtained from the same production batch, were synthesized using methanol as the solvent and had a flat surface with no meniscus because of surface treatment in the wet-gel synthesis process. The other aerogel was produced by the conventional method using \textit{N,N}-dimethylformamide (DMF) as the solvent at Chiba University within the Belle detector upgrade program. Our conventional method for producing aerogels with high transparency is basically the same as that of Panasonic, and it is described in detail in Ref. \cite{cite1}. We measured the optical properties (i.e. refractive index and transmission length) of the aerogels at KEK. The specifications of each aerogel are summarized in Table \ref{table:table1}. The average refractive index was measured by the laser Fraunhofer method \cite{cite1} at the four corners of each aerogel tile. The maximum differences in the refractive index at the four corners were 0.0002, 0.0003 and 0.0002 for the Panasonic1, Panasonic2 and Chiba aerogel tiles, respectively, with an accuracy of $\pm $0.0004. The density of each aerogel was measured gravimetrically.

\subsection{X-ray fluorescence analysis}
To determine the X-ray mass absorption coefficient of silica aerogels, we performed XRF elemental analyses. The coefficient is important for calculating the absolute density of aerogels. A total of eight aerogels representative of those produced by our method were analyzed by an XRF spectrometer (ZSX100e, Rigaku). The aerogels, which had various refractive indices, were synthesized using DMF, methanol, or ethanol as the solvent in the wet-gel synthesis process. Because our hydrophobic aerogels contain trimethylsiloxy groups [--OSi(CH$_3$)$_3$] added to the silica particles, we selectively detected carbon in addition to silica (silicon and oxygen) in the XRF analyses. The XRF spectrometer cannot detect hydrogen, but the effect of this element can be ignored because its mass absorption coefficient value is negligible. The XRF analyses revealed that, on average, the elemental fractions by weight of silicon, oxygen and carbon in the aerogels were 44.6\%, 50.0\% and 5.4\%, respectively. The fractions by weight of the atomic elements were corrected by analyzing quartz, considering that the elemental ratio of silicon to oxygen is 1:2 for quartz. We calculated the mean X-ray mass absorption coefficients for 8.04 keV X-ray photons to be 33.9, 33.5 and 33.1 cm$^2$/g for the aerogels synthesized using DMF, methanol and ethanol, respectively. The experimental error was estimated at $\pm $0.3 cm$^2$/g. The coefficients did not exhibit positional dependence within the margin of error. Therefore, we consider these coefficients to be constant everywhere in the aerogel tiles.

\begin{figure}[t] 
\centering 
\includegraphics[width=0.50\textwidth,keepaspectratio]{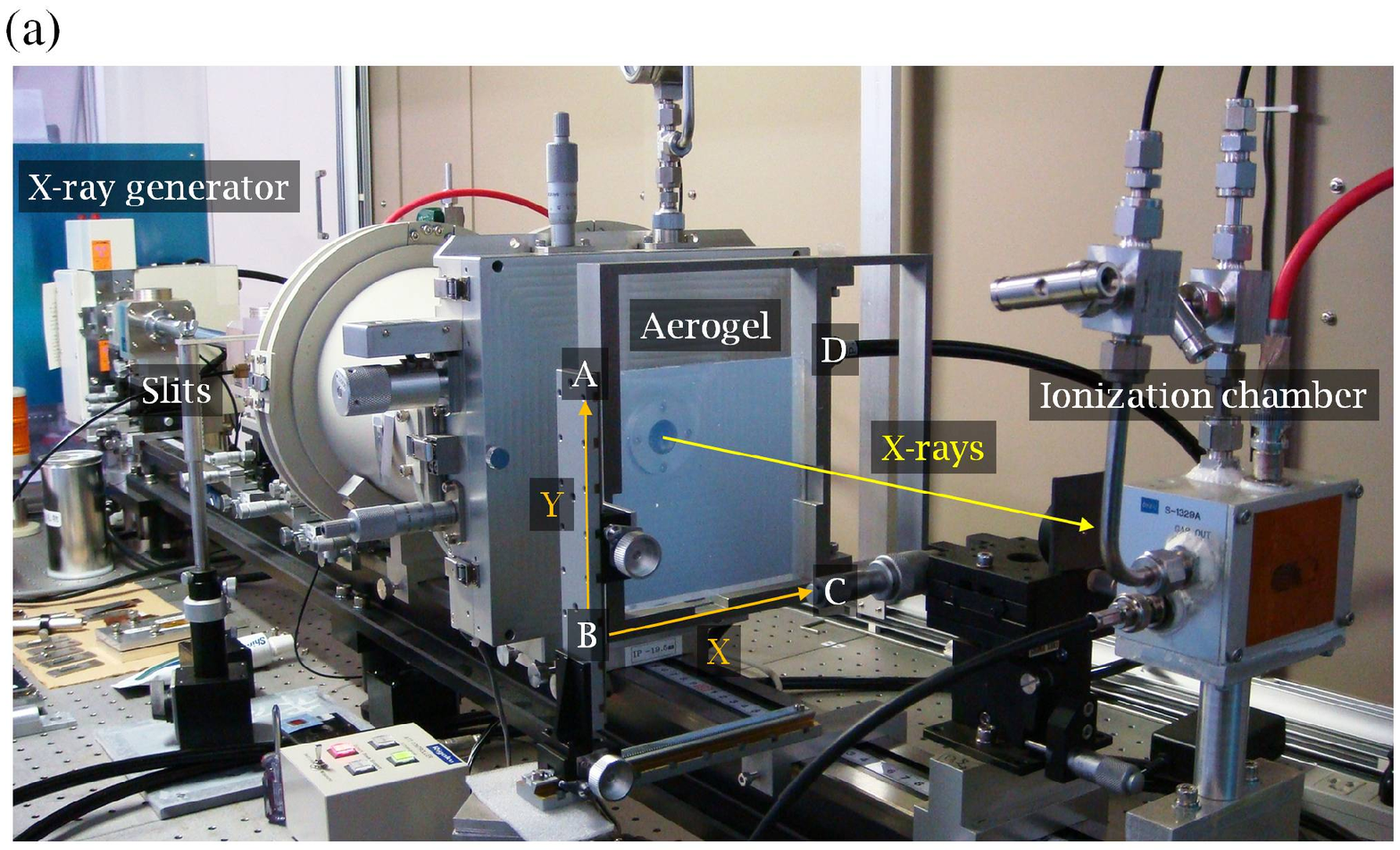}
\includegraphics[width=0.50\textwidth,keepaspectratio]{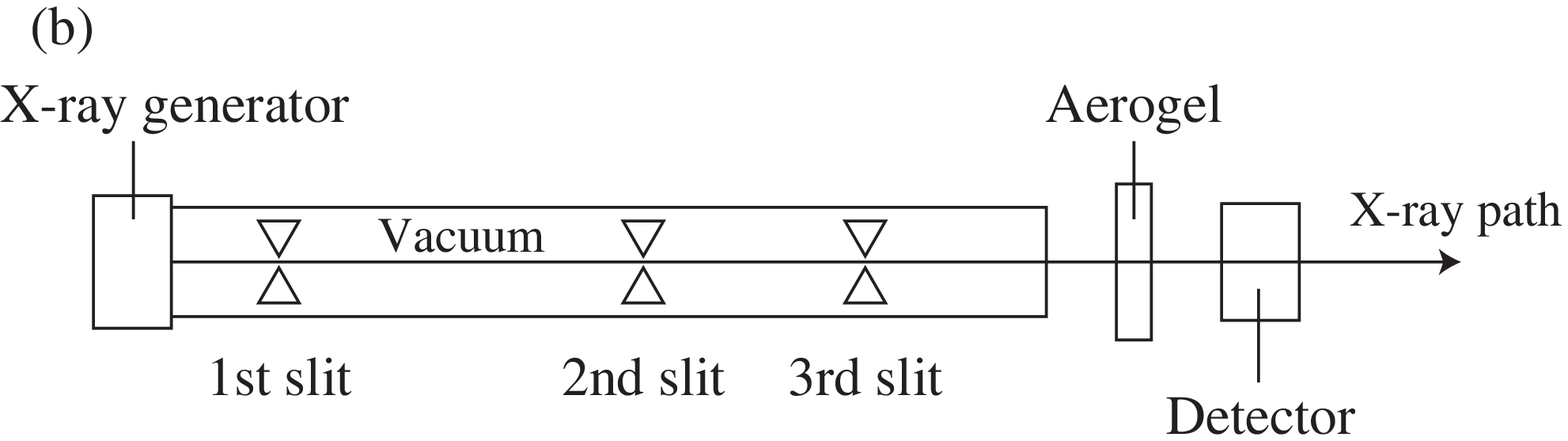}
\caption{Experimental setup of X-ray absorption measurements. (a) Photograph of the setup using the ionization chamber as an X-ray detector. (b) Schematic view.}
\label{fig:fig1}
\end{figure}

\begin{figure}[t]
\centering 
\includegraphics[width=0.27\textwidth,keepaspectratio]{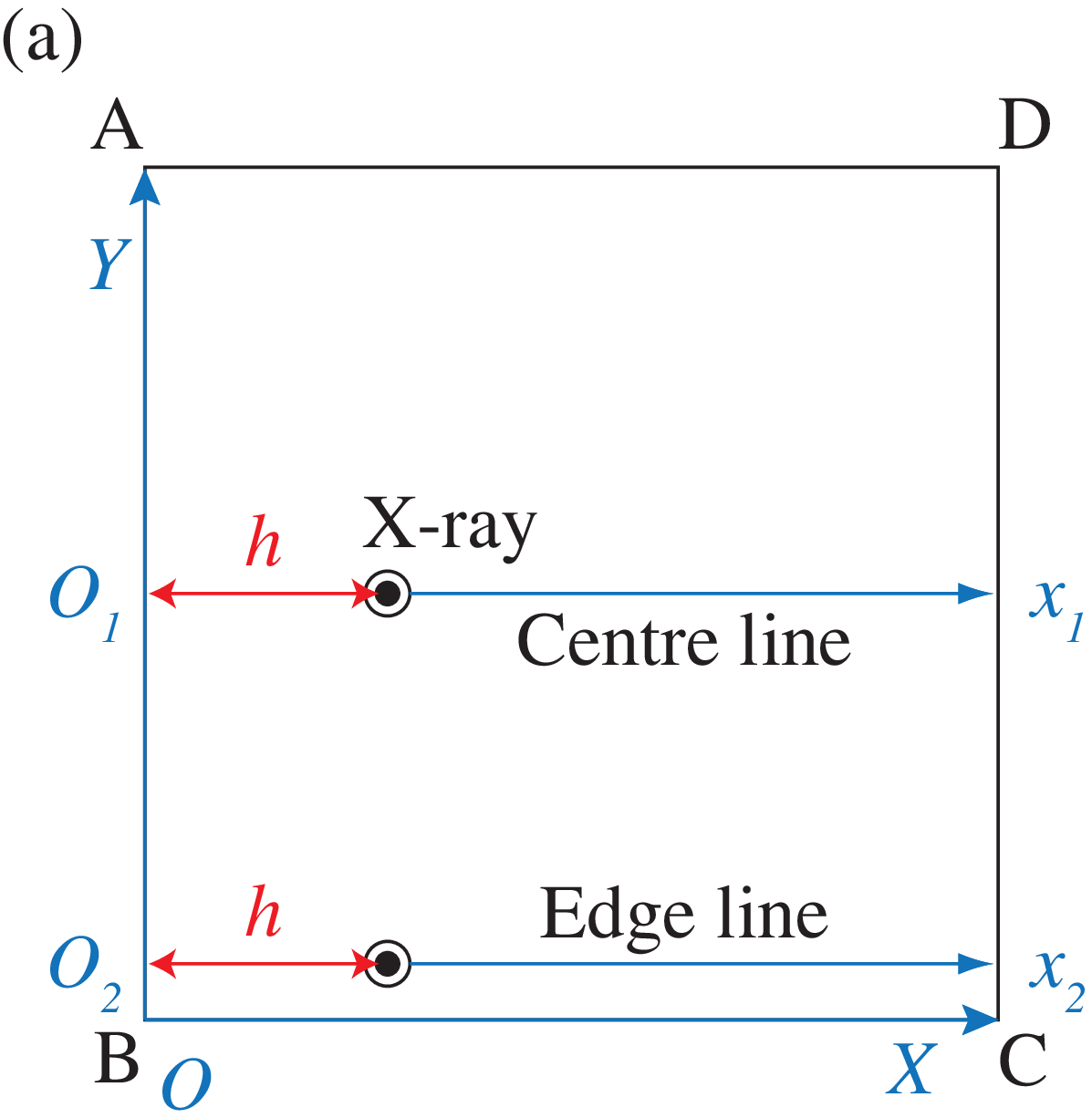}
\includegraphics[width=0.27\textwidth,keepaspectratio]{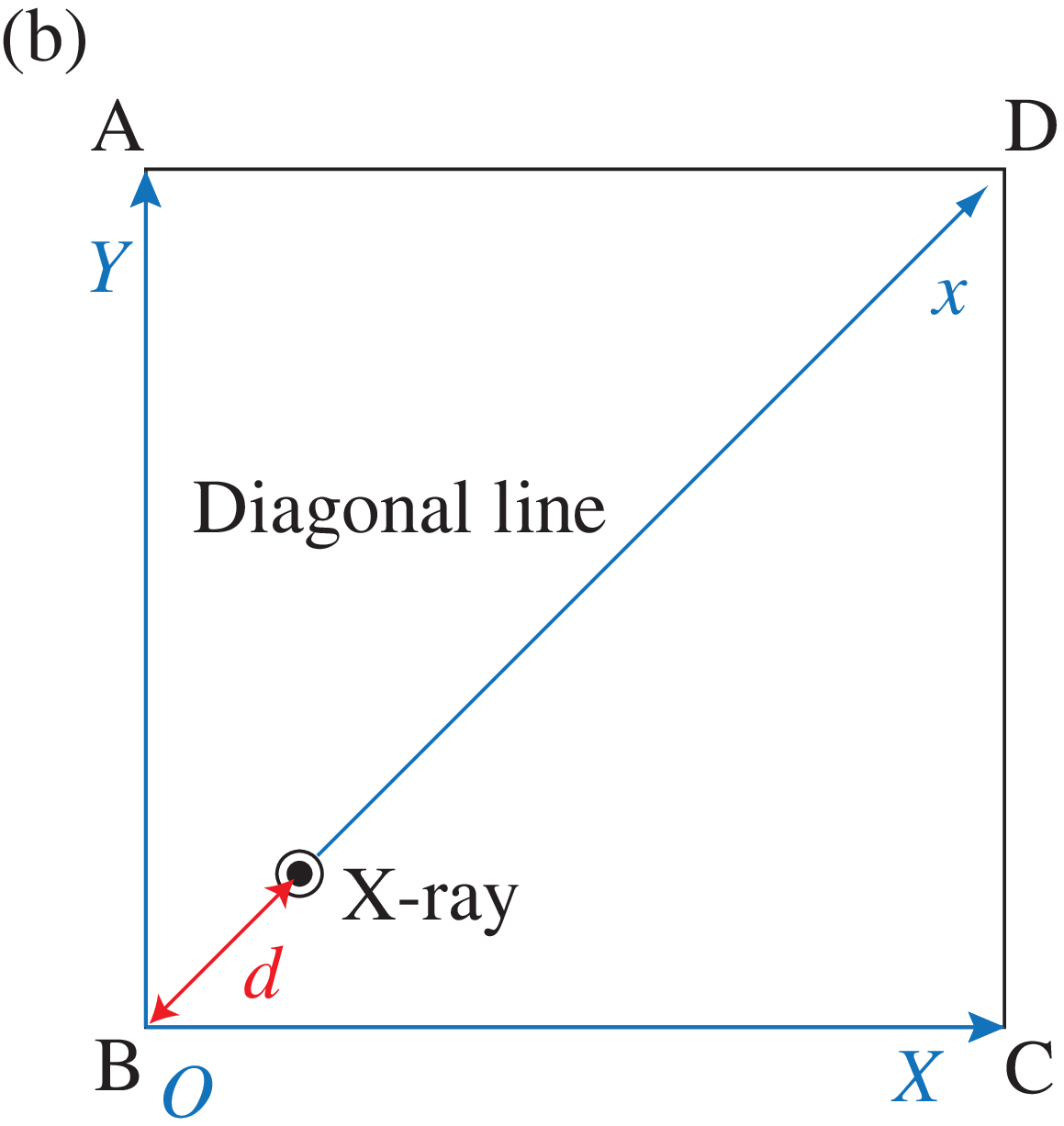}
\caption{Geometrical coordinates on the face of the aerogel tiles from a downstream view. The vertices $A$, $B$, $C$ and $D$ were defined in the counterclockwise direction. With the vertex $B$ as the origin $O$, $X$--$Y$ coordinates were defined for the movable aerogel holder. (a) For the Panasonic aerogels, X-ray scans were performed along the centre ($x_1$ axis, $Y$ = 60.0 mm) and edge ($x_2$ axis, $Y$ = 5.0 mm) lines. The symbol $h$ shows the minimum distance from side $AB$ to the X-ray impact positions. (b) For the Chiba aerogel, an X-ray scan was performed along the diagonal line ($x$ axis). The symbol $d$ shows the distance from the vertex $B$ to the X-ray impact positions.}
\label{fig:fig2}
\end{figure}

\subsection{X-ray absorption measurement}
X-ray absorption measurements were conducted on a NANO-Viewer (Rigaku) system, which consists of a monoenergetic X-ray generator, a focusing incident optic with three slits and two types of X-ray detectors (Fig. \ref{fig:fig1}). The anode target of the X-ray tube of the generator is copper, and its characteristic emission line K$_\alpha $ (with an energy $E$ = 8.04 keV) was used. See Appendix A for more details on the X-ray source. One of the X-ray detectors was an ionization chamber (S-1329A, Oken), whose current was read out using a picoammeter (6485, Keithley Instruments). The ionization chamber has advantages in that we can take data easily and monitor fluctuations in the generated X-ray intensity by installing a subionization chamber upstream of an aerogel sample. However, when the transmitted X-ray intensity was low because the X-rays penetrated thick aerogels, we could not use the ionization chamber. Instead we used, as other X-ray detector, a scintillation counter equipped with a photomultiplier tube (PMT) (R9647, Hamamatsu Photonics). The PMT signals were read out by a combination of a setting rate meter (model 2, Rigaku) and a dual counter/timer (E-541, NAIG); that is, the number of photons detected with the scintillation detector was counted as the X-ray intensity. We prepared an aerogel holder with a movable $X$--$Y$ stage. The holder consists of two one-sided slits on the $X$ and $Y$ rims, respectively, to align the X-ray path with the measurement position on the aerogel tile to an accuracy of 0.5 mm. Using the slits, we confirmed that the X-rays were focused on the holder within a diameter of 1 mm. The meniscus (top) side of the aerogels was turned downstream in the aerogel holder on the X-ray beam line. When we defined a square aerogel $ABCD$ in the counterclockwise direction from downstream (Fig. 2), we placed corner $B$ at ($X$, $Y$) = (0, 0), as shown in Fig. \ref{fig:fig1}a. Except for the space where the aerogel was installed, the X-ray path was in vacuum ($<$100 Pa) in order to avoid scattering of X-rays by air.

The Panasonic aerogels were measured using the ionization chambers. The main ionization chamber was installed downstream of the aerogel for detecting transmitted X-rays; at the same time, the subchamber was installed upstream of the aerogel for monitoring the fluctuation in the generated X-rays. The X-ray scan was performed along the centre line ($x_1$ axis in Fig. \ref{fig:fig2}a, $Y$ = 60.0 mm) and the edge line ($x_2$ axis in Fig. \ref{fig:fig2}a, $Y$ = 5.0 mm) on the aerogels. At each measurement position $h$ (distance from side $AB$ in Fig. 2a) on the aerogels, the currents induced in the ionization chambers by generated and transmitted X-rays [$i_{in}(t)$ and $i_{tr}(t)$, respectively] were recorded 10 times or more every 1 s, where $t = 0, 1, \cdots$ is the time. The current due to the transmitted X-rays was corrected according to the fluctuation in the generated X-rays as follows. The mean current of the generated X-rays $\overline{i_{in}}$ was calculated in each line scan, and correction factors $r(t) = i_{in}(t)/\overline{i_{in}}$ were calculated at each time instant. The corrected transmitted X-ray current $i'_{tr}(t)$ was obtained by dividing $i_{tr}(t)$ by $r(t)$ at each time. Then, at each measurement position $h$, the mean of $i'_{tr}$ over 10 times or more was calculated as $I(h)$. To determine the incident X-ray intensity $I_0$, the X-ray intensities without the aerogels were measured before and after each line scan. To investigate the presence of a possible background, we examined data over approximately 300 s looking for noise in the detector when the X-rays were off. Measurable noise was observed in 22\% of the data, and its mean current was 1.5\% of the typical signal current. The signal-to-noise (S/N) ratio was 10$^3$, so the noise at this level can be considered negligible.

The Chiba aerogel was measured using the scintillation counter. The X-ray scan was performed along a diagonal line (between corners $B$ and $D$ in Fig. \ref{fig:fig2}b) on the aerogel. The accumulated X-ray counts for 5 s were recorded five times to keep the statistical errors (standard deviation of the mean) below 1\% at each measurement position $d$ (distance from the corner $B$ in Fig. \ref{fig:fig2}b). The times of the measurements were also recorded. Then, the mean count of transmitted X-rays was calculated as $I$ at each position. To evaluate the incident X-ray intensity $I_0$, the X-ray counts without the aerogel were measured before and after the diagonal scan. It was known that once the X-ray intensity generated by the NANO-Viewer was stabilized, it decreased monotonically. To estimate the incident X-ray intensities for each measurement time, the X-ray counts without the aerogel were interpolated. During measurement of the incident X-rays without the aerogel, an attenuator (thin aluminum and copper plates) was attached to the surface of the scintillation detector to protect the PMT from the direct X-ray beam. The effect of X-ray absorption by the attenuator was considered in the analysis procedure. To account for possible noise in the detector when the X-rays were off, 5 s background counts (S/N $\sim $ 10$^3$) were subtracted from $I$ and $I_0$.

For both Panasonic and Chiba aerogels, the standard deviations of $I_0$ and $I$ at each measurement position were considered the experimental errors.


\begin{figure}[t]
\centering 
\includegraphics[width=0.45\textwidth,keepaspectratio]{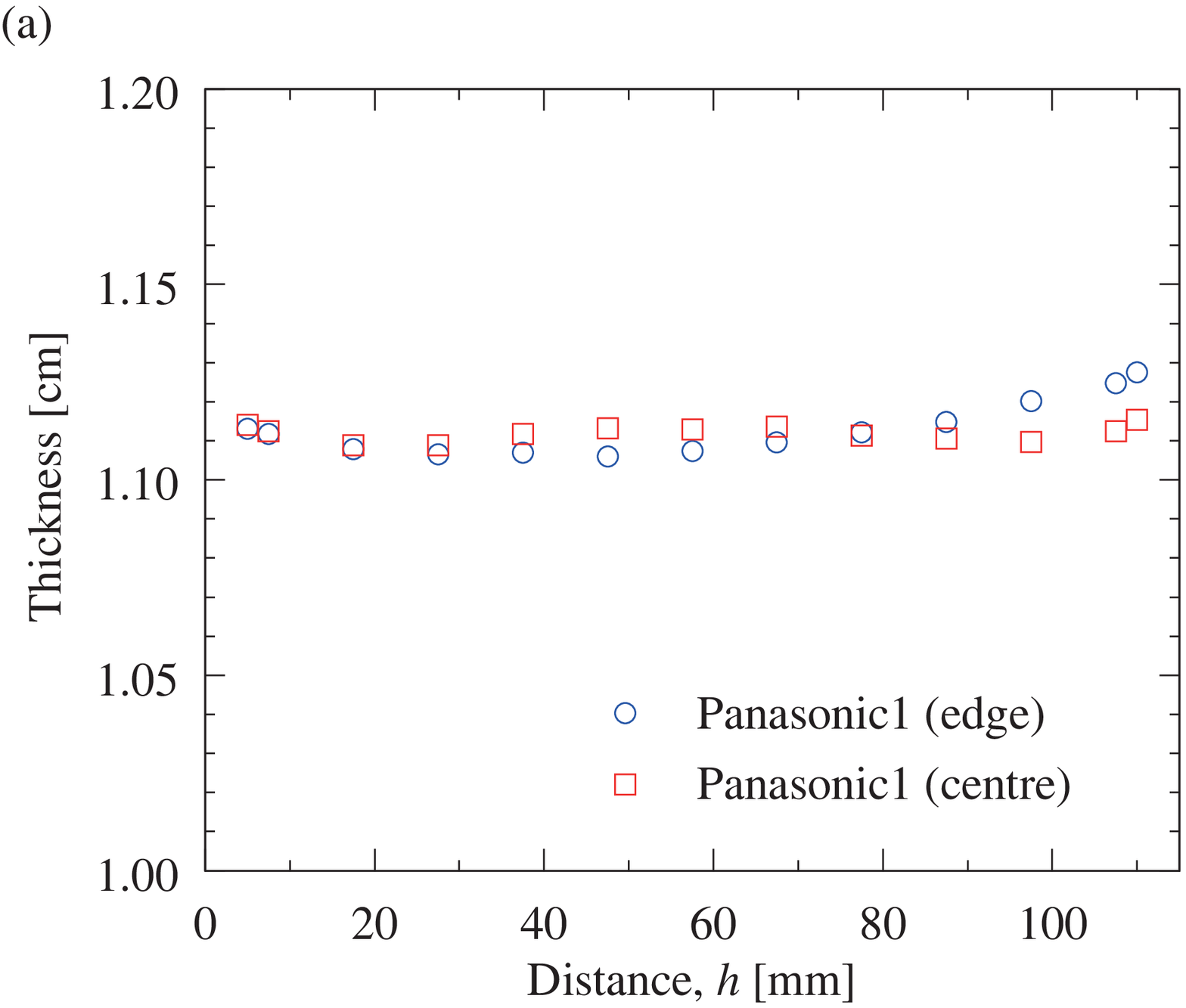}
\includegraphics[width=0.45\textwidth,keepaspectratio]{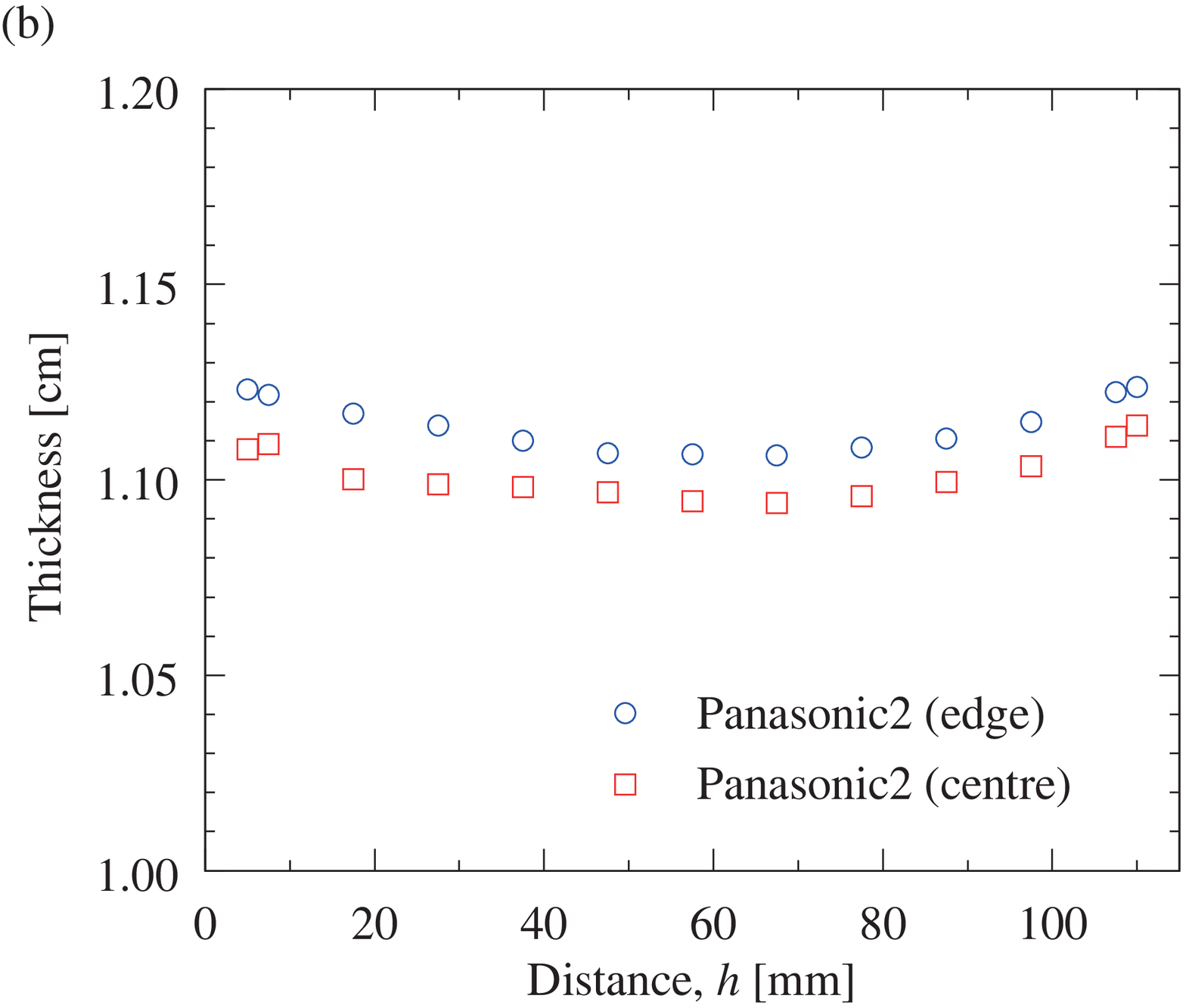}
\includegraphics[width=0.45\textwidth,keepaspectratio]{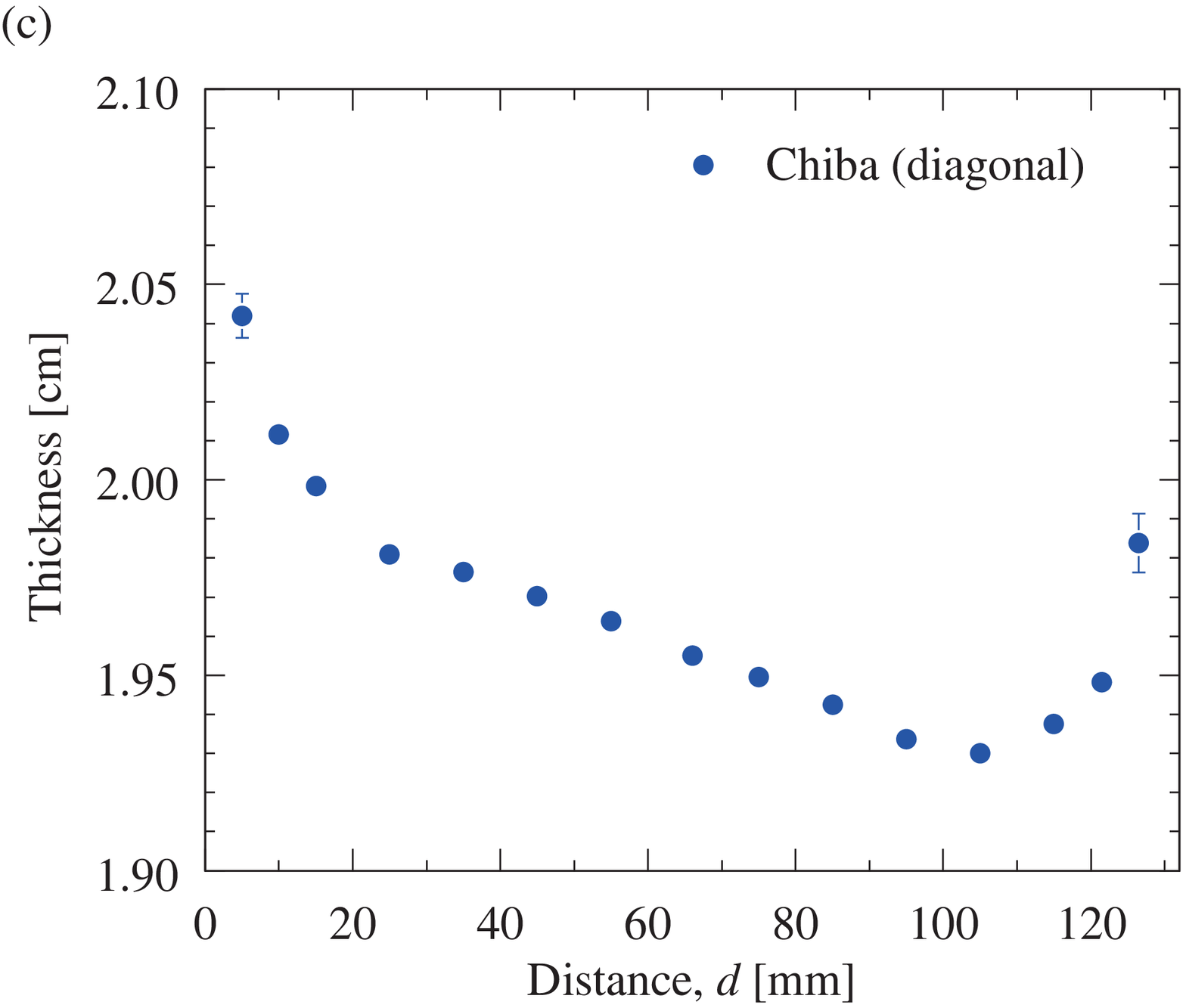}
\caption{Thickness as a function of the distance $h$ from the side of (a) the Panasonic1 aerogel and (b) the Panasonic2 aerogel. The thicknesses measured along the edge line ($Y$ = 5.0 mm) and centre line ($Y$ = 60.0 mm) are indicated by open circles and squares, respectively. The $x$ axis corresponds to the aerogel width. (c) Thickness as a function of the distance $d$ from the corner of the Chiba aerogel. The $x$ axis corresponds to the diagonal line $BD$ of the aerogel.}
\label{fig:fig3}
\end{figure}

\subsection{Thickness measurement}
To evaluate the thickness as accurately as possible, we cut the aerogels along the X-ray scan lines. The Panasonic aerogels were cut with a razor blade, whereas the Chiba aerogel was cut with a water jet cutter. We found that hydrophobic aerogels can be cut appropriately with a water jet cutter. The cross-section of the aerogels was measured with a universal measuring microscope (UMM200, Tsugami), which consists of a fixed microscope and a movable $x$--$y$ stage with a high-resolution (1 $\mu $m) digital position reversible counter (Nikon). It is essential that the position at which the thickness is measured is in line with the position where the X-ray absorption was measured. Considering the size of the X-ray beam and the accuracy of the X-ray incident positions and aerogel cutting, we evaluated the thickness at positions $\pm $0.5 mm from the centre position penetrated by X-rays as well as at this position. The standard deviation of the thicknesses measured at the three positions was regarded as the measurement error at the position where X-ray absorption was measured.

Fig. \ref{fig:fig3} shows thickness as a function of the side distance $h$ (Panasonic) or the diagonal distance $d$ (Chiba) of the three aerogels. The measurement errors ($|\delta t| \lesssim $ 20 $\mu $m) do not exceed the symbol sizes except for the first and last points in Fig. \ref{fig:fig3}c. For the aerogel with no surface treatment, we found that the thickness changed dynamically, particularly at the tile corner, owing to its meniscus geometry.

\subsection{Density evaluation}
\begin{figure}[t]
\centering 
\includegraphics[width=0.45\textwidth,keepaspectratio]{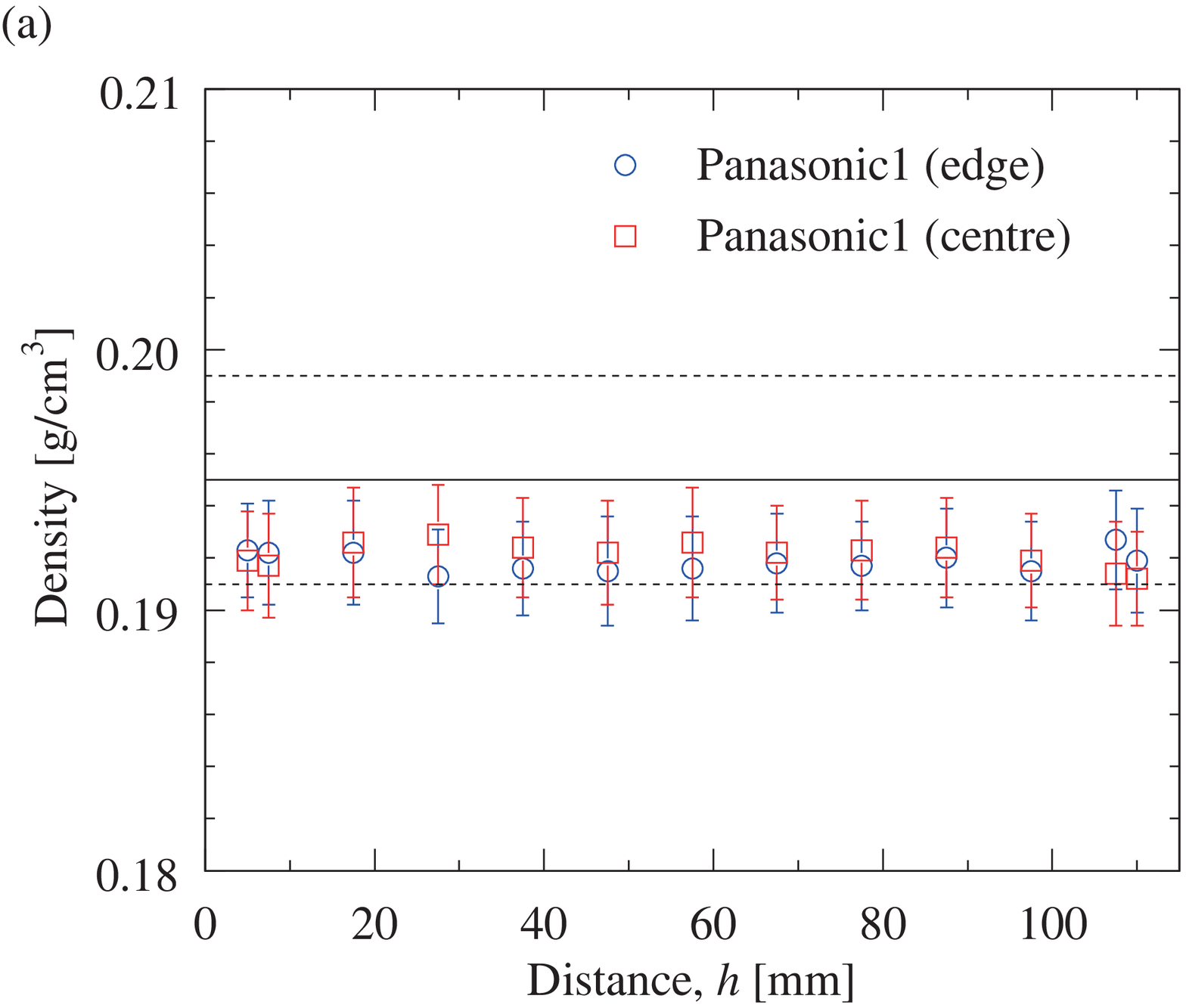}
\includegraphics[width=0.45\textwidth,keepaspectratio]{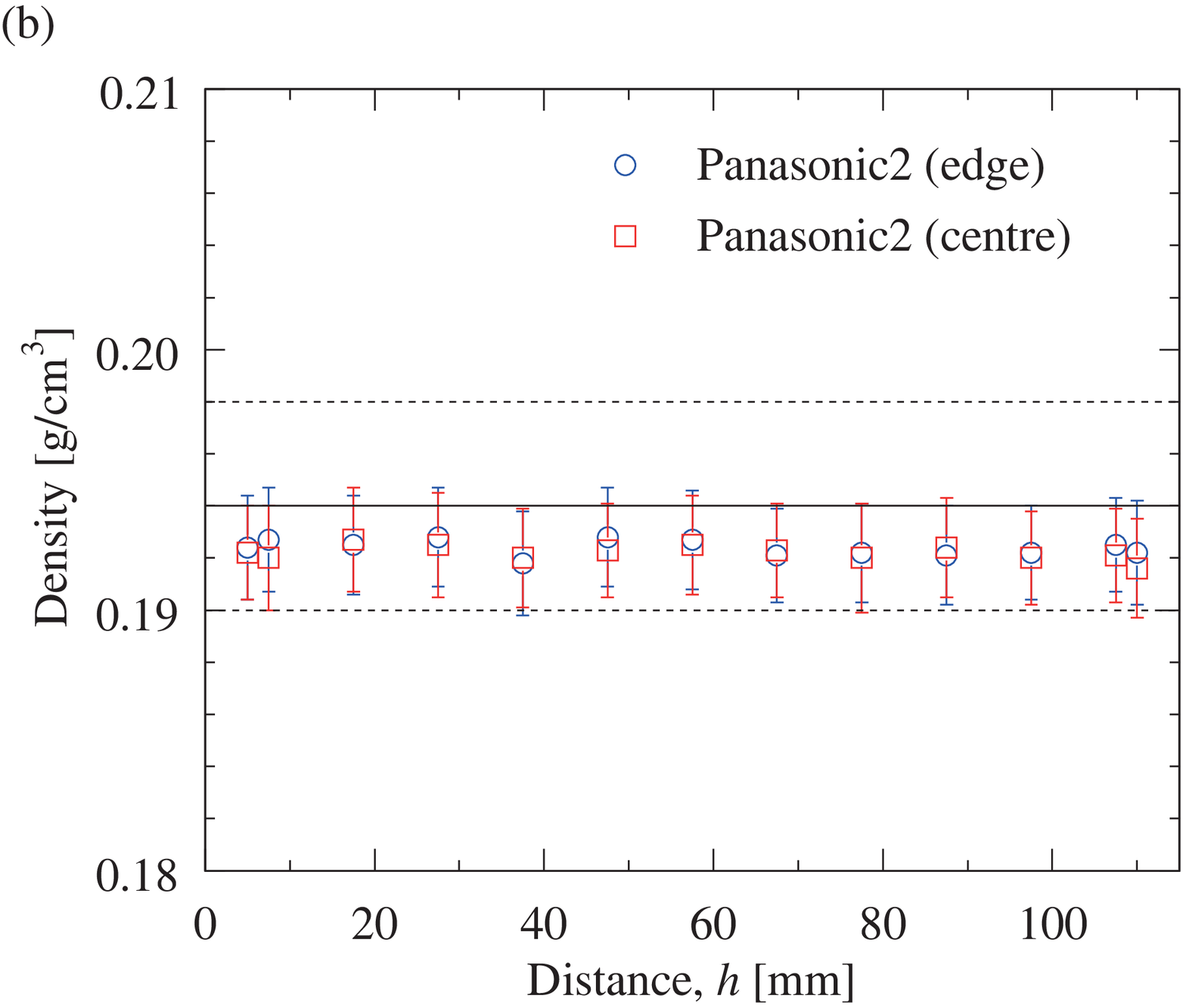}
\includegraphics[width=0.45\textwidth,keepaspectratio]{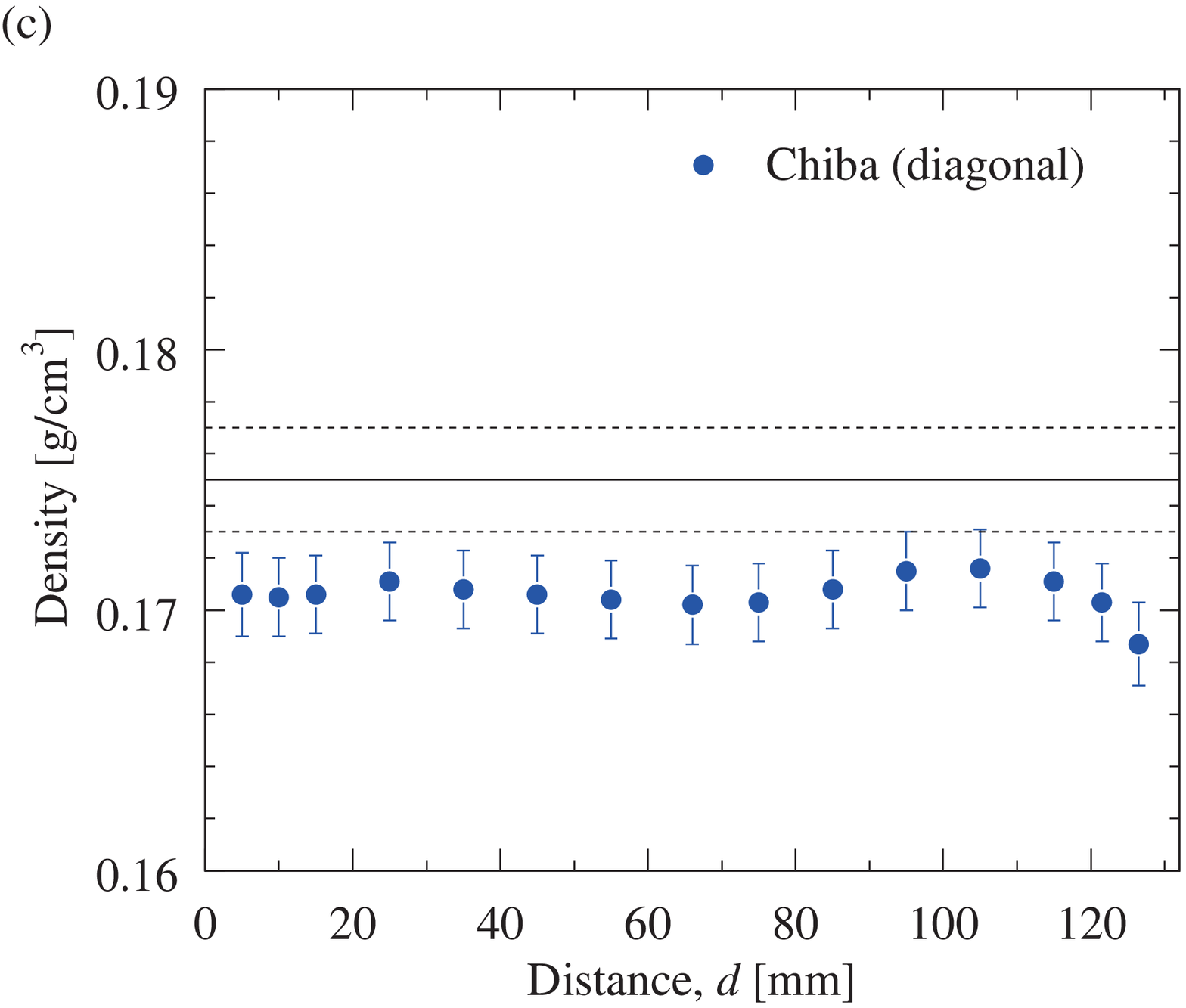}
\caption{Density as a function of the distance $h$ from the side of (a) the Panasonic1 aerogel and (b) the Panasonic2 aerogel. The densities measured along the edge line ($Y$ = 5.0 mm) and centre line ($Y$ = 60.0 mm) are indicated by open circles and squares, respectively. The $x$ axis corresponds to the aerogel width. Solid horizontal line indicates the mean density measured gravimetrically, and dashed lines indicate the range of measurement errors. (c) Density as a function of the distance $d$ from the corner of the Chiba aerogel. The $x$ axis corresponds to the diagonal line $BD$ of the aerogel.}
\label{fig:fig4}
\end{figure}

Finally, the density of the aerogels can be obtained by dividing the mass thickness [$x=(1/\mu _m )\cdot \ln(I_0/I)$] by the thickness at each measurement position $h$ or $d$. Fig. \ref{fig:fig4} shows density as a function of the distance from the side (or corner) of the aerogels. The solid horizontal lines on each plot indicate the mean density measured gravimetrically, and the dashed lines indicate the range of measurement error. The average density over each tile was roughly consistent with the values determined gravimetrically in view of the experimental errors. For each aerogel, the density variation was shown to be within $\pm $1\% relative to the density at the tile centre.

\section{Conclusion}
\label{}
An X-ray radiographic technique was applied to measure the density uniformity of silica aerogel for use as a Cherenkov radiator. The method was described in detail and can be conceived of as a suitable probe for evaluating the density uniformity within an individual aerogel monolith. Panasonic and Chiba aerogel samples were investigated by this method. The experimental results revealed that the density uniformity level in the transverse plane direction was $\pm $1\% for both Panasonic and Chiba aerogels from a trial production. As a further step, uniformity in the thickness direction needs to be investigated in more depth.

\clearpage

\section*{Acknowledgments}
\label{}
The authors are grateful to Dr. H. Yokogawa of Panasonic Electric Works for offering the aerogel samples and to Prof. H. Kawai of Chiba University and his master's students for their assistance in aerogel production. We are also grateful to Prof. K. Bessho of the Applied Research Laboratory at KEK for his assistance in XRF analysis. Thickness measurement of aerogels with the measuring microscope and water jet cutter was performed at the Mechanical Engineering Center at KEK; we are thankful to Engineer M. Iwai for his support. Finally, we thank the members of the Belle II A-RICH group for fruitful discussions on aerogel development. This study was partially supported by a Grant-in-Aid for JSPS Fellows (No. 07J02691 for M.T.) from the Japan Society for the Promotion of Science (JSPS). This publication was supported in part by the Space Plasma Laboratory at ISAS, JAXA.


\appendix

\section{X-rays in the NANO-Viewer}
\label{}
In this appendix, we discuss the monoenergetic X-ray source and the focusing of the X-ray beam in the NANO-Viewer system. The system was constructed for small-angle X-ray scattering (SAXS) studies \cite{cite25} of materials. The simultaneous determination \cite{cite26} of the absorption factor during the measurement of SAXS intensities of a sample is indispensable to such studies. In the present study, we made good use of our experience with X-ray absorption measurements during the SAXS studies.

\begin{figure}[t]
\centering 
\includegraphics[width=0.45\textwidth,keepaspectratio]{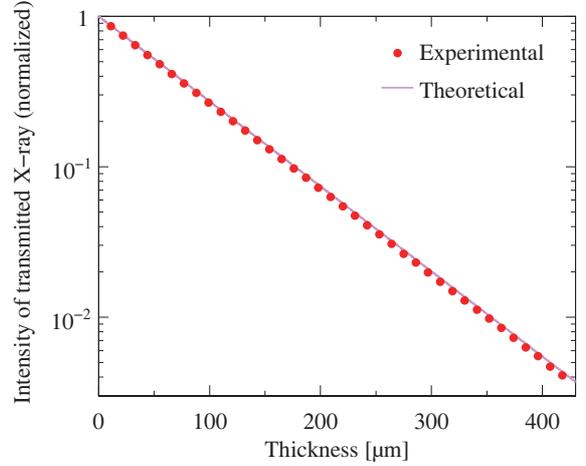}
\caption{Intensity of transmitted X-rays as a function of aluminum foil thickness. Circles and line indicate experimental data and a theoretical estimation, respectively. Aluminum foils with a thickness of 11 $\mu $m were added serially. The intensity of the transmitted X-rays was normalized by the X-ray intensity in the absence of the aluminum foils.}
\label{fig:fig5}
\end{figure}

Because a multilayer mirror is used in the X-ray generator, monoenergetic X-rays are available in the NANO-Viewer system. Fig. \ref{fig:fig5} conclusively demonstrates the monoenergetic hypothesis. This plot was obtained by measuring the intensity of transmitted X-rays while adding thin aluminum foils (11 $\mu $m thick) as a sample material. The intensity of the transmitted X-rays was normalized by the X-ray intensity in the absence of the aluminum foils. The experimental data agree very well with the theoretical estimation.

\begin{figure}[t]
\centering 
\includegraphics[width=0.50\textwidth,keepaspectratio]{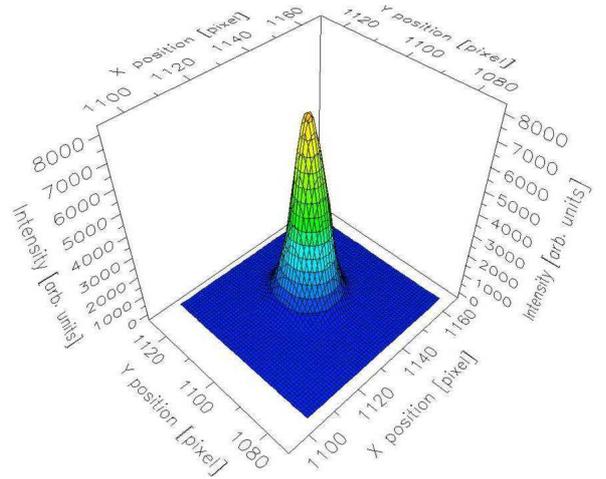}
\caption{Three-dimensional plot of the direct X-ray beam spot in the NANO-Viewer system. The $z$ axis indicates the intensity of the direct X-ray beam. The $x$ and $y$ axes indicate the positions in pixels on the imaging plate. The pixel size was 50 $\times $ 50 $\mu $m$^2$.}
\label{fig:fig6}
\end{figure}

Fig. \ref{fig:fig6} shows a three-dimensional plot of the direct X-ray beam spot. This plot was obtained by analyzing a two-dimensional imaging plate detecting X-ray photons. The positional resolution (pixel size) of the imaging plate was 50 $\mu $m, and Fig. \ref{fig:fig6} suggests that the X-ray spot was distributed over 20 $\times $ 20 pixels. Therefore, we found that the X-ray spot was focused well within a 1 mm diameter by the slits.

\clearpage



\bibliographystyle{model1-num-names}



\end{document}